# Ultrafast High-pressure AC Electro-osmotic Pumps for Portable Biomedical Microfluidics

Chien-Chih Huang[1,2,3,4], Martin Z. Bazant[1,2,3], and Todd Thorsen[1,4,*]

1. Institute for Soldier Nanotechnologies, Massachusetts Institute of Technology, Cambridge, MA 02139 USA

2. Department of Mathematics, Massachusetts Institute of Technology, Cambridge, MA 02139 USA

3. Department of Chemical Engineering, Massachusetts Institute of Technology, Cambridge, MA 02139 USA

4. Department of Mechanical Engineering, Massachusetts Institute of Technology, Cambridge, MA 02139 USA

*To whom correspondence should be addressed: thorsen@mit.edu

This manuscript contains ~3400 words and 6 figures.

Graphical Contents Entry:

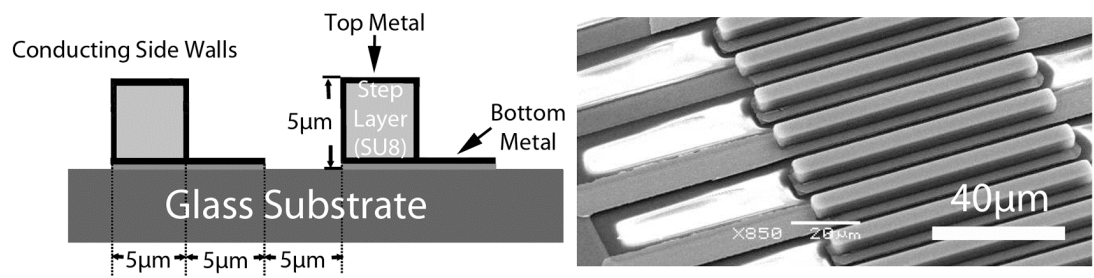

In this paper, the development of a novel low power (mW), low current (mA) AC microfluidic pump is described that is capable of generating kPa head pressures.

Abstract

This paper details the development of an integrated AC electro-osmotic (ACEO) microfluidic pump for dilute electrolytes consisting of a long serpentine microchannel lined with three dimensional (3D) stepped electrode arrays. Using low AC voltage (1 Volt rms, 1 kHz), power (5 mW) and current (3.5 mA) in water, the pump is capable of generating a 1.4 kPa head pressure, a 100-fold increase over prior ACEO pumps, and a 1.37 mm/sec effective slip velocity over the electrodes without flow reversal. The integrated ACEO pump can utilize low ionic strength solutions such as distilled water as the working solution to pump physiological strength (100 mM) biological solutions in separate microfluidic devices, with potential applications in portable or implantable biomedical microfluidic devices. As a proof-of-concept experiment, the use of the ACEO pumps for DNA hybridization in a microfluidic microarray is demonstrated.

# Introduction

Many current analytical techniques in biotechnology can benefit from miniaturization, with the potential to not only minimize costs through reduced reagent consumptions and assay multiplexing, but also though protocol automation. Over the past decade, microfluidic devices have been developed for a large number of molecular and cell biology applications, including the detection of airborne toxins, DNA sequencing, and cell manipulation and sorting[1,2]. While advances in engineering at the chip level have been remarkable, one of the main obstacles to widespread adoption of microfluidic technologies in the clinical and biomedical communities is the need for bulky, expensive hardware to carry out fundamental on-chip processes like fluid manipulation[3].

Despite extensive efforts in micropump development[4], there is still a need for a low power, simple and inexpensive micropump that fulfills the demands for "lab-on-a-chip" devices that require precision manipulation of nanoliters of fluid. The standard approach to microfluidic flow control is based on pressurized actuation of elastomeric membranes[5], which works well in the lab, but requires bulky external plumbing. Over the past decade, DC electro-osmotic porous-plug pumps, capable of high operating pressures (> 10 atm), have been integrated into microfluidic devices[6] and applied to portable fuel cells[7], but the need for high voltage (>10 V) and Faradaic reactions (requiring gas management systems) to sustain DC current can pose problems for biomedical applications and limit miniaturization.

AC electro-osmosis[8] (ACEO) and related induced-charge electrokinetic phenomena[9] are attractive to exploit in integrated micropumps due to the low required voltages (< 3V) and tunable micron-scale flow control without moving parts, although technological challenges still remain. Ajdari[10] first proposed ACEO pumps based on asymmetric inter-digitated electrode arrays, which were realized with planar electrodes of different widths and gaps[11]. State-of-the-art planar ACEO pumps are capable of 0.1 mm/sec velocities in dilute electrolytes with a few Volts, but exhibit flow reversal at high frequency[12], which may be related to ion crowding in the double layer[13]. Traveling wave electro-osmotic pumps yield similar performance at low voltage, and flow reversal at high voltage for all frequencies, which may be due to Faradaic reactions[14]. Very recently, "ultrafast" (> 1 mm/sec) biased ACEO pumps have been reported[15], which involve a DC bias (and thus Faradaic reactions) and larger applied voltages (> 5 $V_{rms}$), closer to those of DC pumps, but with local flow control. Similar velocities have also been predicted at lower voltages (without Faradaic reactions) for optimized three-dimensional (3D) ACEO pumps[16,17], consistent with preliminary experiments by our groups[18,19]. In spite of gains in velocity, however, existing AC electrokinetic pumps still have very low operating pressures of order 10 Pa (0.01% atm) and remain limited to dilute electrolytes (< 10 mM) well below physiological salt concentrations (> 100 mM), an apparently universal feature of ICEO, which may reflect fundamental physical constraints[20-23].

In this paper, we describe dramatic improvements in ACEO pumps for potential integration in portable, biomedical diagnostic devices, achieving much larger pressures (> 1% atm) and higher velocities (> 1 mm/sec) than conventional designs while maintaining a low applied voltage (< 1.5 Volt amplitude). Although the pump uses water as a working fluid, its increased pressure allows indirect manipulation of biological fluids. These designs are shown to be robust against flow reversal and unpredictable operating frequency dependence, which are present in conventional ACEO designs. With a higher possible pressure head, new loading and sample treatment strategies may be employed to avert problems associated with high ionic strength solutions. DNA hybridization in a microfluidic device was performed as a conceptual assay to demonstrate the applicability of 3D ACEO pumps for point-of-care diagnostics. Bridging the gap from theoretical concepts to working systems with such engineering approaches opens avenues for developing biological analysis systems that were previously impractical with ACEO pumps.

## Methods

**Theory.** Until recently, Ajdari's general principle of ACEO pumping by broken spatial symmetry in an electrode array[10] had only been implemented in a planar design[11,12] consisting of flat electrodes of different gaps and widths. As shown in Fig 1(A), this design is inherently inefficient, as portions of the electrode surfaces generate strong local reverse flows, which hinder the dominant forward flow. The same is true of any design that involves small perturbations in height or surface properties around a flat surface, as in Ajdari's original calculations.

Bazant and Ben predicted that much faster flows should be possible (for a given voltage and minimum feature size) simply by changing the shape of the electrodes in three dimensions (3D ACEO), in order to create a "fluid conveyor belt"[16]. The basic idea, sketched in Fig 1(B), is to raise one side of each electrode driving the desired forward flow, while recessing the other side, so that the reverse flow recirculates to form a recessed vortex, which aids, rather than hinders, the forward flow. They predicted at least a twenty-fold improvement over the planar design, and thus ultrafast mm/sec velocities with only a few volts and a 5μm minimum feature size. Urbanski et al. demonstrated significantly improved performance of 3D versus planar ACEO pumps[18,19]**Error! Bookmark not defined.**, but did not test the fastest theoretical geometries.

The design of the next generation ACEO pump presented herein is based on a simple, nearly optimal geometry, recently determined by Burch and Bazant[17.] A schematic of the realized pump design is shown in Fig 2. Each 3D electrode is 10 μm wide with a metal step of width of 5 μm and height of 5 μm. The spacing between electrodes is 5 μm. Coincidentally, these numbers, which form the ratios 1:1:1:1 (bottom exposed electrode width: step height: top electrode width: electrode gap), are close to the theoretical optimum geometry for homogeneous "plated" electrodes[17]. Based on the theory, we expect to observe flows breaking the 1 mm/sec barrier in water (as a standard test fluid) with only a few volts.

Although fast flows are possible, existing ACEO pumps can only generate small maximum pressures ($P_{max}$= 10-100, Pa = 0.01-0.1% atm) against an external load. To understand how this limitation can be overcome, we consider scaling arguments. For any pump operating in the viscous regime of low-Reynolds number, the flow rate decays linearly with the back pressure $P$ according to $\frac{Q}{Q_{max}} = 1 - \frac{P}{P_{max}}$, where $Q_{max}$ is the flow rate at $P=0$. By linearity, the slip-driven ACEO flow in the forward direction is superimposed with a pressure-driven parabolic Poiseuille flow profile in the opposite direction. The situation can be modeled by an equivalent electrical circuit shown in Figure 2(A), where the pump consists of constant current $Q_{max}$ in parallel with the back-flow resistance $R_B$, in series with the external load $R_L$ producing a back pressure $P$. The flow rate $Q_{max}$ is achieved by a closed microfluidic loop fully occupied by the pump (Fig 2(B)).

To estimate these quantities, consider a microchannel of rectangular cross section with a wide floor of width $W$ containing the electrode array and a much smaller height $H << W$ above the upper electrode surface. The flow rate due to a mean slip velocity $U$ over the electrodes is $Q_{max} = \frac{\alpha HWU}{2}$, where $\alpha \sim 1 - (1/2)(H/W)^2$ in the limit $H<<W$. The pressure required to stop the flow is given by

$$P_{max} = R_B Q_{max} = \frac{UL}{k} = \frac{6\eta\alpha UL}{H^2}\left(1 + \left(\frac{H}{W}\right)^2\right) \qquad (1)$$

where $R_B$ is the hydraulic resistance, $k$ is the permeability, and $L$ is the length of the microchannel. According to Eq. (1), the hydraulic resistance to back flow, and thus $P_{max}$, can be increased by reducing the micro-channel height $H$. This strategy has been used to boost the pressure of DCEO pumps by employing linear electro-osmotic flows in submicron pores[6]. For ACEO pumps, the channel height cannot be reduced nearly as much, since it contains micro-fabricated electrodes. Although geometrical confinement can enhance planar ACEO flow by amplifying local electric fields[24], simulations of 3D ACEO show that the channel height (above the steps) should be 2-3 times the step height[17].

On the other hand, the ability to control the flow direction in ACEO gives us another means to boost the pressure, which is not possible with DCEO pumps: increasing the channel length $L$ by using long serpentine channels. Using standard PDMS channels on a single microfluidic chip, the pump length, and thus the pressure, can be easily boosted by two orders of magnitude, which can be further increased by multi-layer pumps connected in series. The resulting pressures exceeding 1% atm open the possibility of indirect manipulation of biological fluids by integrated ACEO pumps in fully miniaturized, portable microfluidic devices, as we shall demonstrate via a DNA microarray.

**Device Fabrication.** Using the aforementioned electrode design, the pump consists of straight 3D ACEO arrays, each ~5 mm long and consisting of 167 electrode pairs. The total area of chip is 2 cm x 3 cm, with a composite of 36 3D ACEO arrays (6012 pairs of the electrodes) on the device. The straight 3D ACEO electrodes are series connected via a serpentine microfluidic channel of height 20 μm ($H$ = 15 μm above the electrode steps), which caps the top of a glass substrate.

The 3D ACEO pumps are fabricated using three layers of photolithography for etching, metal deposition and lift-off processes. First, a 50 nm adhesion layer of chrome, followed by a 50 nm layer of gold is deposited by e-beam evaporation onto a 4 inch borosilicate glass wafer that has been previously cleaned in a piranha solution (1:3 solution of hydrogen peroxide and sulfuric acid) for 15 min. The bottom metal electrodes are created by etching patterns with gold and chrome etchants, defined using standard positive resist photolithography (OCG 825) with high resolution chrome mask and photo-aligner (EVG EV-620). The three-dimensional structures are built up by the second photolithographic process. The 5 μm thick negative photoresist (SU-8) is spin coated on top of the first layer electrode patterned wafer. A second mask is aligned over the first photolithographic features and three dimension steps are formed on top of the bottom electrode after UV light exposure and subsequent solvent development (PM Acetate). The third photolithography step is used to pattern a thick positive photoresist (8 μm, AZ4620) on the same wafer for metal coating. The wafer is tilted in the vacuum e-beam evaporator to cover the whole SU-8 steps with the gold (Fig. 3(A)), followed by a lift-off process with acetone to remove the excess metal and positive photoresist. Individual rectangular devices are die-sawed, and, prior to testing, bare copper leads are attached using conductive silver epoxy. The resulting heights of the three dimensional structures are measured by a white light interferometer (Zygo Corp.).

To assemble the packaged ACEO test device, the base glass substrate containing patterned electrodes is capped with a molded polydimethylsiloxane (PDMS) (Dow Corning- Sylgard 184) microfluidic device fabricated via multilayer soft lithography[5]. To fabricate the PDMS part, two molds were used to fabricate the "injection" and "flow" layers of the device. The injection layer mold consists of a silicon wafer photolithographically patterned with a channel network consisting of positive photoresist (AZ 4620, Clariant). The photoresist structure of the injection layer mold is re-flowed after development (120 °C, 5 minutes, on a hotplate) to create rounded microchannel cross sections. The flow layer mold contains channels patterned in negative photoresist (SU-8 10, Microchem), which has a rectangular cross section. Prior to casting PDMS on the molds, they are treated with trichloromethylsilane (Sigma Aldrich) in vapor phase for 60 seconds to promote mold release. Following silanization, 5:1 part A:B PDMS is cast on the injection layer mold to a thickness of ~ 3 mm while 20:1 A:B PDMS is spin coated (2000 rpm, 30 s) over the flow layer mold. After curing both layers for 80 °C for 20 minutes, the devices are peeled from the mold, and fluidic connections in the injection layer are punched with a flat luer stub adapter. The injection layer parts are subsequently washed with isopropyl alcohol to remove punch residue, dried under $N_2$, and aligned over the channels on the spin coated flow layer. After a secondary bake for one hour at 80 °C to bond the two layers, the composite devices are removed from the flow mold, and interconnects between the top and bottom fluid layer are created by small cuts with scalpel. The ACEO substrate and PDMS cap are then sealed by using plasma treatment. The alignment between substrate and PDMS cap are performed under a dissecting microscope (MODEL) such that the 100 μm wide microchannels of rectangular cross section completely enclose the 80 μm wide 3D ACEO pump structures.

A representative pump and microfluidic device is shown in Fig. 4. The testing loop is in the bottom flow layer, consisting of rectangular cross section microchannels (100 μm (w) x 25 μm (h)), and contains a 180 mm long microchannel section covered by 3D ACEO electrodes and 72 mm of microchannel without electrodes. Channels in the upper injection layer contain the fluid inlets and outlets are rounded and can be gated pnuematically using "push-up" valves at the intersections with the underlying flow microchannels.

Fluid purging and metering operations are software controlled using micro-solenoids connected to the integrated MSL valves. One input channel is used to load operating fluid, deionized water, in the loop. The second input is used to inject flow markers (1.0 μm diameter fluorescent tracers, 505EX/515EM, Molecular Probes) in the bottom parts of the closed testing microfluidic loop away from the electrodes to prevent interference due to electrophoresis or contamination of the pumps with the particles. A signal generator (Agilent 3320A) is used to operate the ACEO pumps at various AC frequencies (0.5-100 KHz), from 1V to 3V P-P. Movies of the moving flow markers inside the channel are recorded using a camera (Sony XCD-V50 B/W, 640 x 480 pixels) under an inverted fluorescent microscope (Zeiss Axiovert 200M) with 20X objective lenses at each driving frequency and voltage. The focal point of the microscope is set to the centerline of the microchannel to record the fastest moving markers. The pumping loop is purged with fresh operating fluid prior to each measurement to minimize variability between successive test runs. Marker movies are analyzed using an open source particle image velocimetry code (URAPIV) through a MATLAB routine, which has been found to provide velocity measurements with less than 5% absolute error in calibration studies.

**Results and Discussion**

The experimental results of fluid velocity versus frequency at different voltages are shown in Fig 5. The mean velocity observed by PIV in the non-pump section of the loop is converted into two characteristics of the pump itself: $v_{pump} = 2Q_{max}/A$, the effective slip velocity over the electrodes in an equivalent shear flow (where $A$ is the cross-sectional area above the electrodes), and $P_{max}$, the head pressure, defined above. The results demonstrate "ultrafast" flows reaching $v_{pump} = 1.37$ mm/sec and head pressures exceeding $P_{max} > 1\%$ atm (100 times greater than existing ACEO pumps), with only 1.06 V root-mean-square AC voltage (3 Volts peak to peak). Under these conditions, the power consumption is 4.8 mW at current 3.5 mA, so the pump could operate continuously for days with a typical Li-ion battery, thus enabling fully miniaturized portable devices.

Following the simulations of Bazant and Ben[16], planar ACEO pumps with the same minimum feature size (electrodes widths of 5 μm and 30.5 μm, and spacings between electrodes of 5.5 μm and 18.5 μm) were also tested in the same geometry for comparison. As expected, the 3D ACEO pump is much faster for all voltages and frequencies, although by a smaller factor (5x) compared to simulations (20x). Although the maximum flow rate is around 1 kHz for both designs, the 3D ACEO pump has a second peak around 10 kHz, which could be useful in applications to further reduce Faradaic reactions. Moreover, as predicted[16,17], the 3D ACEO pump exhibits robust pumping in the forward direction, while planar pump reverses at high frequency and high voltage. This puzzling feature first reported by Studer et al.[12] has recently been attributed to the crowding of counter-ions in a highly charged double layer[13,20].

Dramatic advances in the cost reduction of genetic testing have been made over the last few years through both increased parallelism and miniaturization, principally through the use of DNA microarrays. Compared to traditional lithographically patterned or mechanically spotted microarrays, which rely on diffusion for the hybridization of DNA probe-complement pairs, microfluidic DNA hybridization assays are much more efficient, utilizing microchannels to flow DNA targets over the surface of complement DNA. Many groups have developed microfluidic molecular hybridization devices, demonstrating pM specific target-complement sensitivities on the order of minutes for the microfluidic platforms vs. tens of hours for traditional diffusion-based, high density DNA microarrays[25-29].

As a proof-of-concept experiment, the 3D ACEO pump, attached to an external microfluidic manifold, was used to specifically hybridize a fluorescently-tagged target DNA oligonucleosides against spotted complement DNA. ACEO pumps were implemented as the sole pressure source, where deionized water in pumping loop was used to pump the high ionic strength DNA hybridization solution (~100 mM) though a second microfluidic chip via Tygon connection tubing (Fig. 6(A)). A straight microchannel was mounted on a glass slide containing spotted DNA probes (both complementary and non-complementary, consisting of 3x repeats of the 20 base pair oligos, Mag1 and Apn1, DNA barcode sequences from the *Saccharomyces* gene deletion project[30]). 20 mer Cy3 fluorescently-labeled DNA target solution complementary to the Mag1 DNA sequence (1 nM in 3x SSC buffer, 0.1% sodium dodecyl sulfate) was subsequently loaded into Tygon tubing at the inlets of the bypass devices and pumped through the devices using the ACEO pump driven at 3V AC for three minutes. A cooled CCD camera (Apogee Alta U2000) was used to capture the fluorescence at each spot. The capture images indicate positive hybridization between the labeled Mag1 sequence and the positive (+) spotted complement, and very low background fluorescence generated by non-specific hybridization between the Cy3-Mag probe and the non-complementary (-) spotted Apn1 probe (Fig. 6(B)). The relative fluorescence intensity ratio (+ vs. – spots) was > 50:1. The absolute positive fluorescence intensity of the positive hybridization was nearly four times greater than the results of similar experiments in the absence of pumping.

Simple scaling arguments based on our experimental results show how to design serpentine 3D ACEO pumps with desired characteristics. The flow rate or pressure can be increased by connecting multiple pumps in parallel or in series, respectively. For example, since our prototype pump consists of only one thin layer (20 μm (h) channels), its pressure can be increased by a factor of ten, exceeding 10% atm simply by stacking ten layers (for a total thickness below 1 mm). Regardless of the channel layout, for a given device volume, there is always trade-off between maximum flow rate and maximum pressure. We have already noted that to maximize pressure, the channel height, *H*, should be reduced as much as possible, given the electrode sizes and fabrication methods, so this should be viewed as a constant when designing the channel layout. To tune the flow rate, we can vary the channel width *W*.

Material and fabrication constraints limit the total device cross-sectional area per channel *A*, which includes the surrounding walls and substrate thickness, and is thus larger than the internal channel cross-sectional area *HW*. The fabrication method thus sets the ratio $\beta = HW/A$. For a given volume *V*, the total length of the channel can be estimated as $L = V/A$, ignoring any corner effects in regions of the channel without a pumping surface. Using (1), we find

$$P_{max} Q_{max} = \beta \gamma \frac{U^2 V}{H^2} \qquad (2)$$

where $\gamma < 1$ is a constant, reflecting the hydraulic resistance of corners and connections, compared to the pumping regions. For fixed velocity, volume, and channel height, Equation (2) shows that the maximum pressure is inversely proportional to the maximum flow rate. It also determines the required volume for the pump, given target specifications of flow rate and pressure for a given application.

## Conclusions

In summary, we have demonstrated dramatic enhancements in driving pressure and flow rate of AC electro-osmotic pumps, both theoretically and experimentally, using optimized 3D electrode shapes and long, serpentine channels. The designs are shown to be robust against flow reversal and can achieve pump velocities over 1 mm/sec and head pressures exceeding 1% atmosphere, while keeping the AC voltage below 1.1 Volt rms. With a higher possible pressure head, the device was used to remotely pump aqueous DNA solutions with moderate to high salt concentrations (>100 mM), illustrating the potential for 3D ACEO pumps to be incorporated as low-power pumping elements in field deployable lab-on-a-chip devices.

## Acknowledgements

This research was supported by the U.S. Army through the Institute for Soldier Nanotechnologies, under Contract DAAD-19-02-0002 with the U.S. Army Research Office.

# Figures

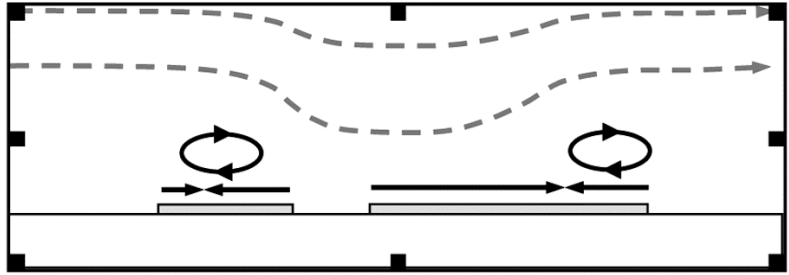

(A) Planar design

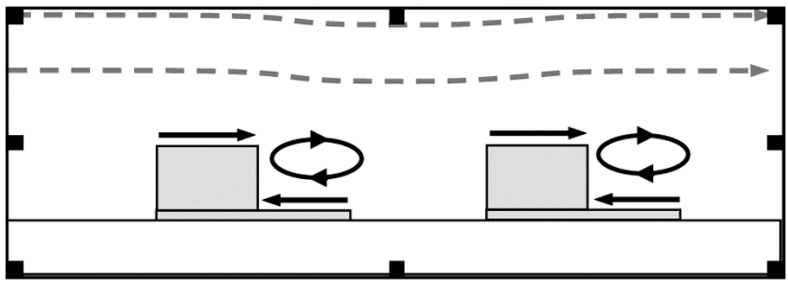

(B) Optimal 3D design

**Figure 1**. Schematic diagram of fluid flow resulting from an ac voltage applied between two electrodes in a periodic array. Surface slip on the electrodes and resulting pumping streamlines are represented with solid and broken lines, respectively. (A) Asymmetric planar electrodes produce directional pumping through a biased competition between opposing slip developed on the electrode pairs. (B) Optimal non-planar 3D electrodes drive recessed counter-rotating vortices to form a much more efficient "fluid conveyor belt".

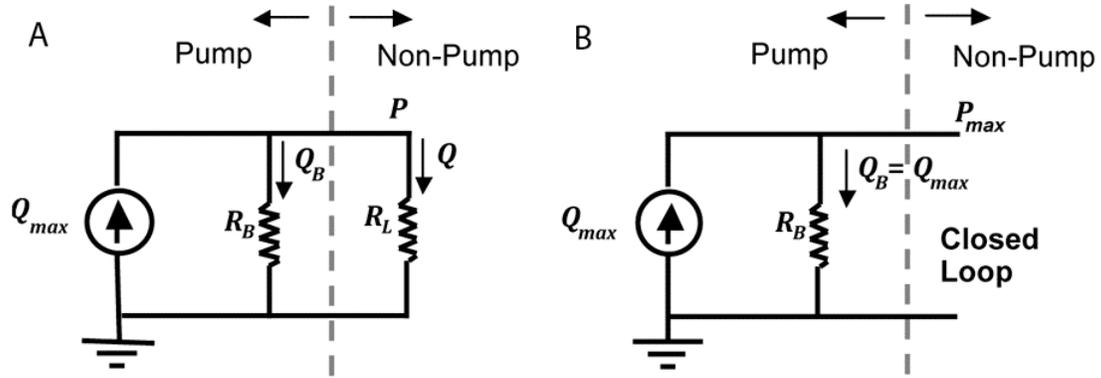

**Figure 2**. Fig 2. Equivalent circuits for a slip-driven pump with internal hydrodynamic resistance $R_B$ and maximum flow rate $Q_{max}$ either (A) in series with an external load $R_L$ producing a back pressure $P$ or (B) in a closed loop with back flow $Q_B = Q_{max}$.

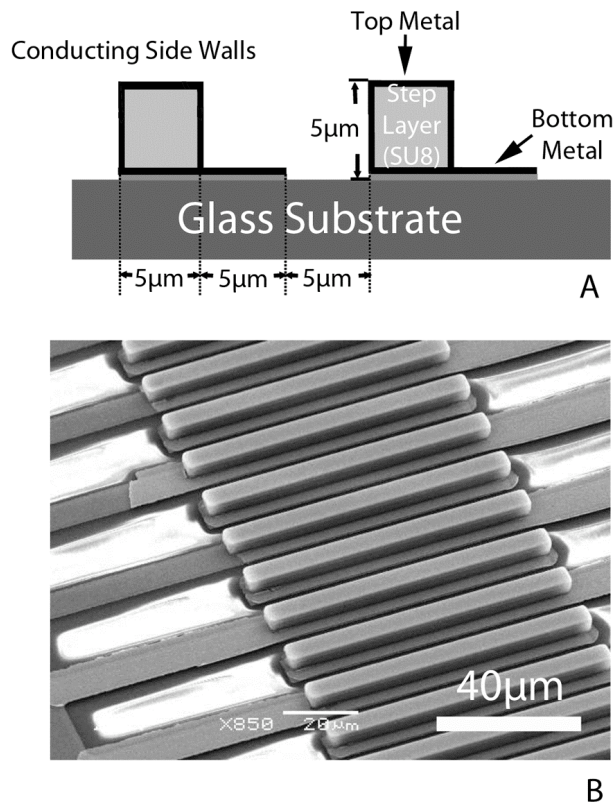

**Figure 3**. A) Detail of three dimensional steps on the interdigitated electrodes. (B) Scanning electron microscope image of a 3D ACEO pump comprised of gold coated electrodes patterned on an insulating substrate.

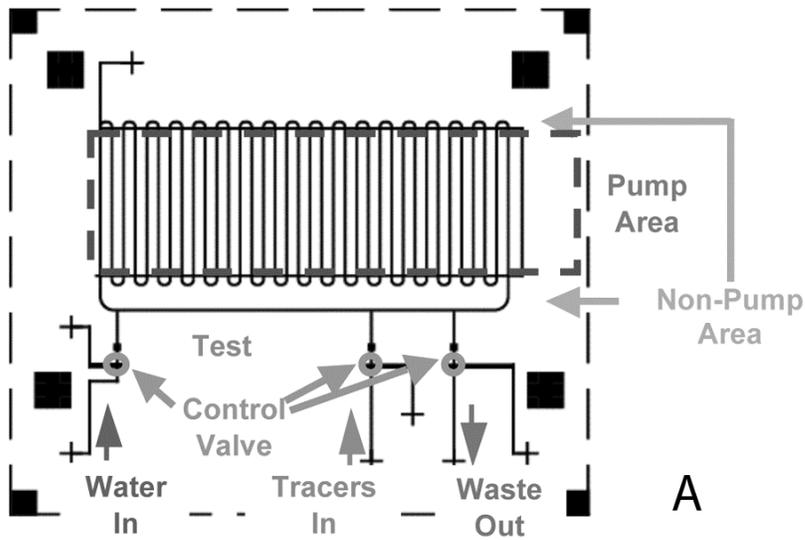
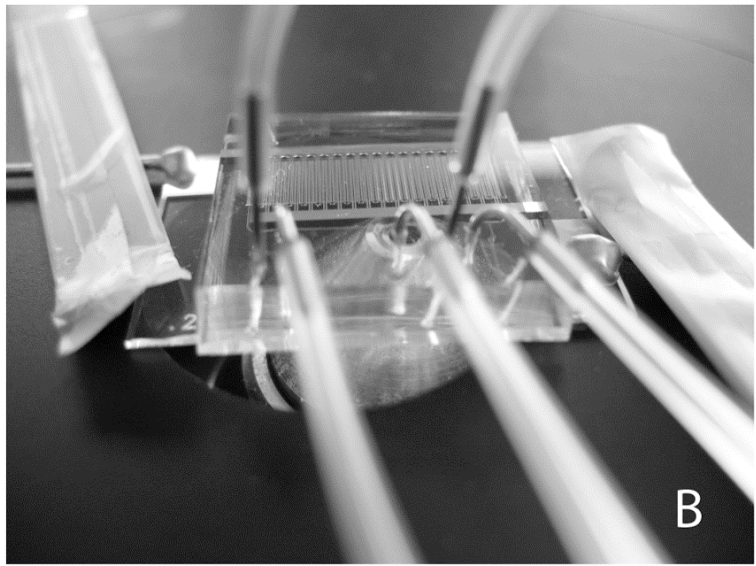

**Figure 4**. (A) The microfluidic device containing an integrated pump loop with rectangular channels which linked 3D ACEO step electrodes in series allows for repeatable characterization. When an ac voltage is applied to the electrodes, the fluid is pumped in clockwise direction within the closed circuit. (B) The photo of the 3D ACEO pump measurement setup.

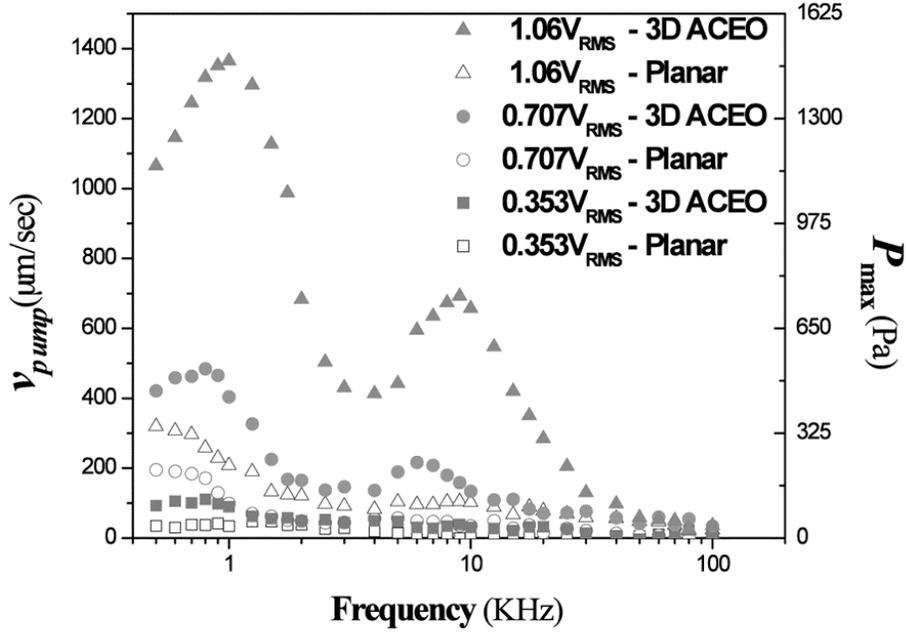

**Figure 5**. Flow rate versus AC frequency for the 3D ACEO pump of Figs. 3 and 4, compared to the fastest planar ACEO pump design in the same geometry, at peak-to-peak voltages 1 V, 2 V and 3 V (corresponding to the root-mean-square voltages indicated). The observed velocity away from the pump is converted to a mean slip velocity over the electrodes $v_{pump}$ and maximum pressure $P_{max}$, which characterize the pump's performance, independent of the hydraulic load in the non-pump section of the loop.

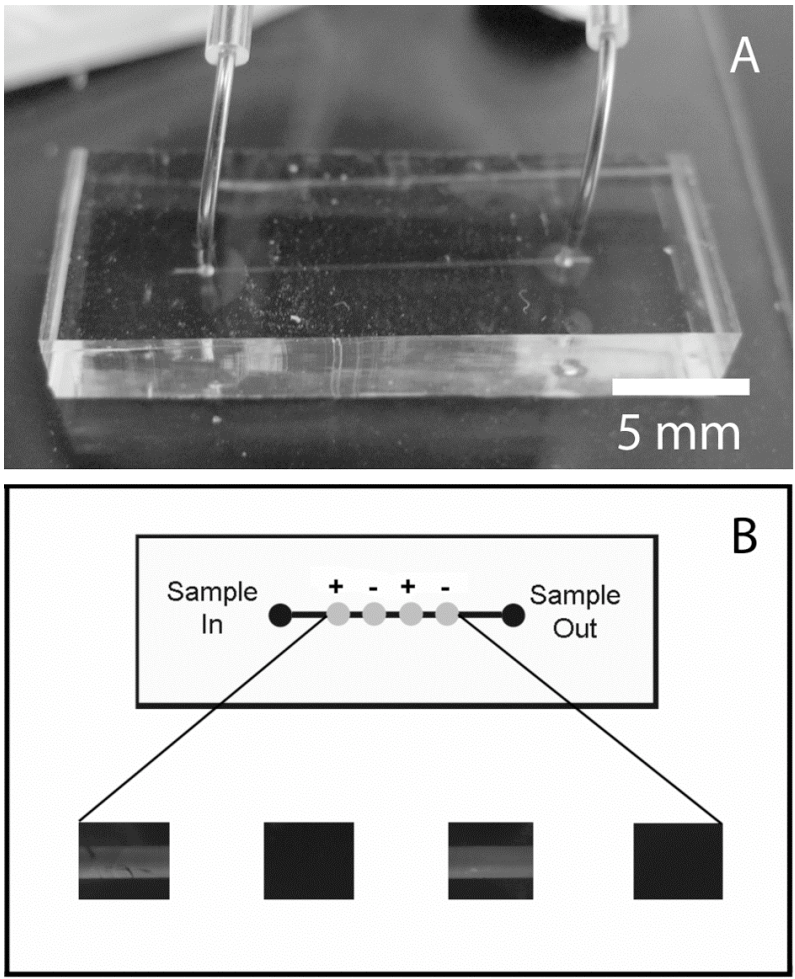

**Figure 6**. (A) Photography of the microchannel where DNA hybridization is performed. The in and out tube is connected to the 3D ACEO pump to execute off chip pumping. (B) The fluorescent result of the DNA hybridization at the complementary Mag1 spotted sites (+) and the non-complementary Apn1 spotted sites (-).